\documentstyle[preprint,aps]{revtex}
\begin{document}
\draft
\title { Structure of binary clusters: are icosahedra relevant? }
\author{ Stefano Cozzini$^{1,2}$ and Marco Ronchetti$^3$}
\address{$^1$ Dipartimento di Fisica, Universit\`a di Trento and INFM,
I-38050 Povo, Italy \\
$^2$ Departamento de Fisica Atomica, Molecular
y Nuclear, Universidad de Sevilla, Apartado 1065, 41080 Sevilla
-Spain \\
 $^3$ Dipartimento di Informatica e Studi Aziendali, Universit\`a
di Trento, 38100 Trento - Italy}
\date{\today}
\maketitle

\begin{abstract}
We investigate the structure of 13-particle clusters in binary alloys
for various size ratios and different concentrations via MD
simulation. Our goal is to predict which systems are likely to form
local icosahedral structures when rapidly supercooled from the melt.
We calculate the energy spectrum of the minimal energy structures, and
characterize all detected minima from both their relative probability
and from a structural point of view. We identify regions in our
parameter space where the icosahedral structure is dominant (like in
the corresponding monoatomic case), regions where the icosahedral
structure disappears and other where icosahedral structures are
present but not dominant. Finally, we compare our results with
simulations reported in literature and performed on extended binary
systems with various size ratios and at different concentration

\end{abstract}

\section {Introduction}
Over the last decade, chemists and solid state physicists have been
greatly interested in the platonic solid with the highest symmetry:
the icosahedron. Chemists became interested in icosahedra because
several clusters have been found to be icosahedral or
poly-icosahedral.  In particular, the recent synthesis and structural
determinations of metal carbonyl clusters resulted in a variety of
icosahedral clusters containing transition metals and main group
elements \cite{pic}. Solid state physicists believed until recently
that icosahedral symmetry played no role in extended structures
because it is not compatible with periodic arrangement of structural
units (i.e. with crystalline order). It is now known that icosahedral
symmetry is compatible with quasiperiodic translational order, and
states of matter arranged in such a way have been found
(i.e. quasicrystals \cite{jar}). Moreover, icosahedral structures are
suggested to be important in disordered systems, like supercooled
liquids and glasses \cite{nel89}. The presence of icosahedra in simple
supercooled liquid and glasses was revealed by means of computer
simulations: from the pioneristic work of Steinhard, Nelson and
Ronchetti\cite{snr81} up to the most recent works \cite{jon95}, there
is evidence in literature that it is so for monoatomic supercooled
liquids and glasses interacting via Lennard-Jones (LJ) and via a
variety of different metallic potentials \cite{rif}. So far however
studies of LJ binary systems yield contrasting results
\cite{ron94}. The study of icosahedral clusters is therefore important
both in itself and because it can give hints on the presence of such
structures in disordered systems and on the nucleation of extended
icosahedral quasicrystals.  Computer simulation performed by Honeycutt
et al.\cite{hon87} on homogeneous LJ systems showed that icosahedrally
symmetric clusters are the lowest in energy up to a size of 5000
atoms. A similar study in binary systems would be interesting, but
unfortunately it is difficult because the parameter space to be
investigated is far more complex. Infact in the case of monoatomic
species only the interaction potential and the cluster size (i.e. the
number of particles) have to be specified, while in binary systems the
relative abundance of the two species, geometric factors (i.e. size
ratios) and parameters relative to the binding energy have to be taken
into account. For this reason computational studies on mixtures are
rare and focused on specific aspects, like the study of impurities in
clusters \cite{gar89a} or the dynamics of phase separation
\cite{cla93}.  To start exploring clusters in a binary system it is
therefore necessary to reduce the search space by fixing a few
parameters and studying the dependence on the remaining ones. For
instance, both the above referred works by Garzon {\it et al.} \cite{gar89a}
and Clarke et al.\cite{cla93} keep the geometric parameters fixed
and vary the energetics. In the present work we decided to cut the
parameter space in an orthogonal direction by fixing the energetic
parameters and the size of the cluster, and varying particle size and
concentration. We therefore use the same depth of the potential well
for both atomic species and for the interaction between unlike
particles. For the cluster size we focus on 13-particles clusters,
since our aim is to determine the importance of icosahedral
structures. We therefore study the geometric structure and energy
spectrum of 13-atom clusters for four different atomic size ratios and
for all possible relative concentrations.  In the next section we
present the details of the computational model. In section III we
discuss the methods of measurements that we used. The results are
presented in section IV, followed by a discussion and a
comparison with the literature (section V).

\section {The Computational Model}
We studied a system composed of 13 particles belonging to two
different species (L and S) and interacting with a LJ isotropic
potential. The two species differ because of geometric factors: the
radius of L-atoms is larger than the radius of S-atoms. The
interaction between S-atoms is characterized by the parameters
$\sigma_{SS}$ and $\epsilon_{SS}$: the interaction potential is
therefore $V(r)=4\epsilon_{SS}[(\sigma_{SS}/r)^{12}-
(\sigma_{SS}/r)^6]$. The potential for the atomic specie L was defined
by $\epsilon_{LL}=\epsilon_{SS}$ and $\sigma_{LL}=\alpha\sigma_{SS}$:
the parameter $\alpha$ therefore fixes the ratio between the radii of
the two species. The interaction between unlike particles was defined
by $\epsilon_{LS}=\epsilon_{SS}=\epsilon_{LL}$ and
$\sigma_{LS}=(\sigma_{LL}+\sigma_{SS})/2$. The values of $\sigma_{SS}$
and $\epsilon_{SS}$ were suited to Argon (this choice does not
invalidate the generality of the results, since its only consequence
is to fix the energy and length scales). The mass is not a relevant
parameter when only structural properties are of interest, and we
therefore used equal masses for the two species (again, we chose Argon
mass). In the following all results are given in reduced units with
$\epsilon_{SS}$ the unit of energy, $\sigma_{SS}$ the unit of lenght
and $(m\sigma_{SS}^2/48\epsilon_{SS})^{1/2}$ the unit of time.\\
The only relevant parameter is therefore $\alpha$ which
fixes the size ratio between the two species. In the limiting case
$\alpha$=1 the system is monoatomic since the parameters for L and
S-atoms coincide. We used four different values for $\alpha$: 1.6,
1.4, 1.33 and 1.25.  A cluster in such model is further identified by
the number N of particles composing it and by the relative
concentration of L and S-atoms. We fixed N=13 and studied all possible
concentrations of $\eta \in
\{1,12\}$, where $\eta$ is the number of S-atoms (the limiting cases
$\eta$=0 and $\eta$=13 are equivalent, since they both corresponds to
a monoatomic system). Our investigation covers therefore 48 points in
parameter space.  Our simulation method was the Molecular Dynamics
technique \cite{ron90} in the microcanonical ensamble. We used Verlet
algorithm with a time step of 0.01 in our reduced units. Clusters were
kept in free boundary conditions, and we checked that no atoms
evaporated during the runs (cases in which atoms evaporated were
discarded since we were interested only in 13-atoms clusters). For
each value of $\alpha$ and $\eta$ we produced a large number of
independent realizations using two different algorithms which yielded
similar results. The first procedure consisted in starting (at
equilibrium) from a liquid monoatomic cluster composed of large
particles. Then $\eta$ large atoms were randomly substituted by small
atoms, and the obtained cluster was allowed to relax for 10000 time
steps so as to equilibrate the new system. The cluster was then
quenched into a minimum of its potential energy hypersurface. The
whole process was repeated for a reasonably large number of times
(typically 1000 times). We collected the final configurations and
examined them by studying the resulting energy spectrum and the
geometric configurations corresponding to the most relevant final
states.  The second procedure consisted in evolving for a very long
time (typically 20 million steps) a liquid cluster formed by $\eta$
small particles and $13- \eta$ large particles, taking snapshots every
10000 steps. The temperature is high enough that two configurations
separated by 10000 time-steps are statistically independent from
each-other. Each of the snapshots was then relaxed until the cluster
reached the local minimum in potential energy, resulting in 1000 to
2000 independent configurations, which were then studied.  Since we
use a microcanonic Molecular Dynamics, by using the second procedure
all the snapshots (for a given concentration) have the same total
energy, while in the former case each initial configuration has
different total energy. In spite of the different procedures, the two
processes yield equivalent results: in particular the energies of the
quenched clusters found with the two methods are the same, suggesting
that the configurations we obtain do not depend on the particular
history of the sample.\\
To reach the minimum energy configuration we
used a steepest descent minimization: the kinetic energy was quickly
stolen so that the cluster fell into the nearest potential energy well
without having many chances to perform additional explorations in the
configuration space while approaching a local energy minimum. The
method corresponds roughly to a cooling rate of 10$^{14}$ K/sec.  For
each of the 48 points in our parameter space we produced at least 2000
configurations. For a set of points a first analysis revealed complex
behaviours: in such cases we collected more statistics, producing up
to 4000 additional configurations.  To avoid that some very rare
configurations gain a considerable weight in our statistic, we consider
only those minima which appear at least twice in our collection.

\section {The Methods Of Measurements}
We were interested in identifying the structures which characterize
the quenched clusters from both a structural and an energetic point of
view. A first classification can be obtained by looking at the energy
distribution of the collection of frozen samples: for each value of
$\alpha$ and $\eta$ we study the energy spectrum, i.e. we count the
number of different potential energies reached at the end of the
quenching process by our samples. There will be a lowest minimum,
corresponding to the absolute minimun energy ( i.e. the lowest minum
in the potential energy surface) and then set of many {\it excited}
states. If there are significant gaps in the spectrum and moreover
different structural properties are detected on either side of the
gap, we can say that different {\it phases} are present in the
system. Here we use the concept of {\it phase} discussed in depth by
Honeycutt and Andersen \cite{hon87}: a phase corresponds to a set of
structures of mininum energy (inherent structures). Different phases
are characterized by different sets of structures with different
properties.\\
  Not all energy minima will be reached with equal
probability: we therefore associate to each minimum the corresponding
frequency of visits, defined as the ratio between the number of times
that a particular value of energy has been reached and the total
number of samples produced. Having obtained this information, we can
plot the frequencies versus energies: such a plot reveals
interesting features of the system under investigation. \\
 We then study
the geometric configuration of the quenched clusters. The structural
analysis has been performed with various methods presented in
literature: Voronoi polyhedra, Common Neighbours Analisys
(CNA)\cite{hon87} and Steinhardt's invariants \cite{snr81}. These (and
other) methods for investigating the presence of icosahedral
structures were recently described in detail in a paper \cite{ron94}
where their validity was investigated.  The technique of the Voronoi
polyhedron is our main instrument. Given a cluster, our algorithm
identifies the central atom(s) (as the atom(s) with the highest number
of first neighbors) and then tries to determine the corresponding
polyhedron(s). Polyhedra classification is based on the number of
faces and on the number of edges of every face. A typical polyhedron
is expressed with a set of 5 indexes indicating the number of faces
with 3,4,5,6,7 edges. (The probability to find faces with more than 7
edges is negligible). In this way we characterize the structure of
every cluster: for instance the polyhedron for a particle at the
center of a perfect icosahedral arrangement is the dual of the
icosahedron, i.e. a dodecaedron (12 faces, each of them with 5 edges):
the set is therefore (0,0,12,0,0). In a few pathological cases the
program was not able to costruct the polyhedron: the cause of the
failures is essentialy due to the fact that for some clusters the
central atom has too few neighbors to allow our code to work (i.e. the
cluster is very elongated). \\
As second tool we used the CNA diagram
in the new formulation recently proposed by J\'onsson \cite{glm94}. We
classify every cluster by looking at the CNA diagrams which involve
the central atom. This technique allows to discriminate between
different configurations which exhibit the same Voronoi polyhedron
(such cases are mostly due to different arrangements of particles on a
second shell). This method also permits to distinguish between
configurations which are geometrically similar but differ for the
relative placement of L and S-atoms.  Finally, we still needed to
distinguish between clusters which were classified as icosahedral by
the above methods, but which had a different degree of distortion. We
found the use of the invariants proposed by Steinhardt et
al.\cite{snr81} very effective for such measure.  A number of clusters
were also examined visually with the help of computer graphics
programs \cite{glm94,mdt}. The number of configurations
inspected in such way is obviously limited when compared to the great
number of clusters produced: we looked mostly at the configurations
identified as interesting by the previous analysis. We found the
results of visual inspection very useful for reaching a better
understanding of the phenomenology.

\section{Results}
We now present the results of our investigation . We divide the
presentation in two steps: first we discuss the energetics of the
clusters by studying the number of minima with respect to the energy
(energy spectrum) and the number of visits ot every minima (frequency
of visits). Later we will examine the corresponding structures to
understand in more detail the behavior of the systems.

\subsection{Energetics}
The energy of the clusters is first examined by looking at the energy
spectrum as a function of $\eta$. Fig.~\ref{fig1} shows the
spectra for the various size ratios ($\alpha$=1.25 in fig.1A,
$\alpha$=1.33 in fig.1B, $\alpha$=1.4 in fig.1C, $\alpha$=1.6 in
fig.1D). In each of the figures an horizontal segment shows an energy
level. On the x-axis we have the different values of the concentration
$\eta$: the number of small atoms grows when moving from left to
right. We show a window having a height of $0.3\epsilon$, which is large
enough to give a complete representation of the energy distribution
above the absolute minimum. The heigth of such window is comparable to
that reported by Honeycutt and Andersen on the simulation performed on
monoatomic 13 atom cluster \cite{hon87}. In some cases this range
includes the region where evaporation begins to occur.  For each
energy level E and for each concentration $\eta$ we then study the
frequency $\phi_{\eta}(E)$, defined as the number of times we find
a configuration having an energy between E and E  + $\delta$E over the
total number of configurations produced. Although $\delta$E is small
($0.01\epsilon$) it gives a coarse graining effect because of the even
smaller energy differences among the minima. Frequencies
$\phi_{\eta}(E)$ versus E are reported in Fig.~\ref{fig2}, where again the
letters A, B, C and D refer to the ratios $\alpha$=1.25, 1.33, 1.4 and
1.6 respectively. In each figure we plot a line for each value of
$\eta$, and in order to increase readability we perform a shift
proportional to $\eta$ of the value $\phi_{\eta}(E)$ (i.e. the
lowest line corresponds to $\eta$=1, the next lowest to $\eta$=2 and
so on). The highest peak at an energy E indicates that the most likely
configuration (for that particular values $\alpha$ and $\eta$) has
energy E. If only energetics was important, the most visited
configuration should always be the one at the lowest energy. We will
see that this is not always the case, showing that also entropy plays
an important role in determining the most relevant state(s). \\
 Having
described the structure of the figures, we now pass to their
interpretation and discussion. The most striking feature of the
spectra is the existence of an energy gap which is most evident for
$\alpha$=1.25: it is however smaller than for a monatomic system
\cite{gar89b}. Below the gap the states are discrete, while above it a
continuum of states exists, showing the signature of a liquid-like
state. The continuum is not completely sampled by our simulations.  At
$\alpha$=1.25 for high values of $\eta$ all the states fall in the
continuum, while for values of $\eta$ smaller than 8 isolated states
are present above the gap but below the continuum. The number of such
isolated states grows as $\eta$ decreases: they tend to populate the
gap for small values of $\eta$. The lowest bound of the continuum
seems to follow a parabolic-like behavior as a function of $\eta$,
reaching a minimum for $\eta$ between 5 and 7. Also the lowest energy
state follows a similar behavior. For $\alpha$=1.33 a similar
behavior is observed, with a few differences. The gap tends to close
for low $\eta$'s due to two effects: the lowest energy moves toward
higher energies and the number of discrete states below the continuum
grows. At $\alpha$=1.40 such tendency is enhanced, leading to a
qualitative transition: it is not clear any more that a gap exists
when $\eta \le5$. At $\alpha$=1.60 such trend is confirmed: the gap
disappears already at $\eta$=7. We will further discuss the the gap in
a later section, after having identified the structures responsible
for this behavior. Figure 3 shows the value of the lowest energy as a
function of $\eta$ for the various values of $\alpha$. Of course the
minimum energy has the same value for all $\alpha$'s when $\eta$=0 or
$\eta$=13, since these cases describe the same system: a monoatomic
cluster. The energy always decreases when including an impurity of
type l ($\eta$=12) and increases if the impurity is a small enough
particle ($\eta=1$, $\alpha \le 1.33$).  As we already mentioned, the
energy behavior is parabolic-like for $\alpha$=1.25, with a minimum
for $\eta$=7. As $\alpha$ reaches the intermediate values
($\alpha$=1.33, $\alpha$=1.4) the shape does not change: the minimum
remains at $\eta$=7 but the energy of the minima increases in a more
pronounced way on the left side of the curve (i.e. in the region where
the concentration of small atoms is smaller). At $\alpha$=1.6 however
the situation is quite different: the curve presents two different
minima respectively at $\eta$=4 and $\eta$=9, implying the existence
of two different mechanisms. Again, before discussing further this
point, we will need to perform some structural examination.\\
The examination of frequency of visits shown in figure 2 enriches the
above observations. For $\alpha$=1.25 we can see that the energy states
below the gap are the most frequented. Above the gap instead the
states in the continuum follow a rather flat distribution, except for
$\eta$=1 and 2 where the lowest isolated states above the gap but below
the continuum show a significant occupancy.  For $\alpha$=1.33 the
scenario is similar, with a notable difference: for $\eta \le 3$ the
states at the lowest energies (below the gap) are not the most visited
ones: rather, the most frequented states are the lowest in energy
above the gap. The same is true for $\alpha$=1.4 and $\eta \le 5$. At
this size ratio it is also evident from figure 2C that the gap
disappears for the lowest half of the range.  The case $\alpha$=1.6
shows an interesting behavior. At high values of $\eta$ the scenario is
similar to that of the previous cases: the dominant peak is at the
lowest energy. When $\eta$ decreases the height of the dominant peak
decreases and the gap shrinks. Above the gap a continuum of states has
a rather flat distribution. The gap disappears for $\eta$=8. For $5 \le
\eta \le 7$ no structure can be observed, as in the low range for
$\alpha$=1.33 and 1.4. At $\eta$=4 however a new, different feature appears:
a peak develops at the lowest energy. The height of this peak grows
when $\eta$ decreases. This time no energy gap separates the high peak
from the structures at higher energies. Finally, the case $\eta$=1 is
singular: the most frequented states are the two at the lowest
energies, where twin peaks are found. The energy of these states is
much higher than for $\eta>1$.\\
  The behaviour of the energy gap is
shown in fig.~\ref{gap}. The gap grows with $\eta$, and converges to a
similar value for all concentrations when only one L particle is
present ($\eta$=12). The presence of a clear gap in a similar
monoatomic system is known to be the signature of the presence of
icosahedra \cite{hoa79}. The value of the gap for monoatomic systems is
$0.219\epsilon$ \cite{gar89b}, close to the value we get for $\eta=12$ for all
$\alpha$'s. As we will see later, when discussing the icosahedral
structures, this seems also to be the case for binary mixtures.

\subsection{Structural Analysis}
Having identified interesting regions and peaks, we analyse the
samples to find which structures are responsible for the behavior we
described above. Due to the very large number of minima found, it is
obviously unfeasable to examine in detail all of the corresponding
clusters. We therefore decided to focus our structural analysis on the
configurations having a percentage of visits greater than 1.5\%. In
addition, we arbitrarily decided to examine also a few other
configurations which we considered to be potentially interesting (like
for instance the lowest energy configurations also in cases where they
are not frequently visited). 3-D images of selected configurations
helped understanding the structure of selected clusters.\\
  A first
indication comes from an analysis with Voronoi
polyhedra. Figures~\ref{fig4} (A,B,C and D for $\alpha$=1.25, 1.33,
1.4 and 1.6) shows the percentage of the most relevant polyhedra as a
function of $\eta$. To avoid cluttering the figures with too many
data, we show here the behavior of two most relevant polyhedra, and
collect all the other polyhedra in a common group. We can see that for
$\alpha$=1.33 and high $\eta$'s the dominant structure is a 12-faced
polyhedron. At about mid-range the presence of such figure starts
dropping, and steadyly decreases down to a 10\% for $\eta$=2. The
growth of the signal of other-than-12-faced polyhedra at low
concentrations is mainly due to 10-faced polyhedra: only 11 atoms form
the Voronoi cage while the two remaining particles are on the second
shell. A such cluster is reported in fig.~\ref{fig5}: the
signature of its cage is (0,2,8,0,0), meaning that 2 faces have 4
edges, 8 faces are pentagonal and no face has 3, 6 or 7 edges.  The
scenario is similar for $\eta$=1.4, the difference being that the drop
of the 12-faced polyhedron is much sharper, and very few such
structures are found in the range $1\le\eta\le4$. Again, the
competing structure is a 10-faced polyhedron which dominates the
low-$\eta$ region.  The behavior is instead quite different in the two
extreme cases: $\alpha$=1.25 and $\alpha$=1.6. In the first case the
12-faced polyhedron is the dominating king over the whole range: other
structure comes close to a noticeability edge only for
$\eta\le2$. Quite the opposite situation is found at $\alpha$=1.6:
12-faced polyhedra survive only for very high $\eta$'s (i.e. for a
cluster of S-atoms with a few L impurities). At the other end of the
range (i.e. in a cluster of L-atoms with few S impurities) the
dominating structure is the 9-faced polyhedron. The 3D graphical
representation of such polyhedron for the case $\eta$=2
(fig.~\ref{fig6}) indicates that the lowest minima are achieved by
clustering the small atoms in the center and accomodating all the
large atoms in an external shell. In the particular case illustrated
in figure the resulting cluster can be seen as two interpenetrating
9-faced polyhedra each one centred on the small atoms. Such trend
seems to be rather general: also figure ~\ref{fig5} shows that
small atoms tend to for a droplet which becomes the core of the
cluster. Both 9-faced and 12-faced polyhedra are of little relevance
in the middle of the region, where no structure is found to dominate:
the peak of other structures is a mixture of unequal, elongated
clusters.  It's worth to note that for all values of $\alpha$ there a
visible presence of 12-faces structures at $\eta$=1, even when the
low-$\eta$ range is characterized by structures other than 12-faced
polyhedra. \\ Of course the analysis with the Voronoi polyhedra does not
tell the whole story. For instance, 12-faced polyhedra can be of
different sorts: they can be icosahedra with  an L particle at the
center, icosahedra with an S-atom at the center, or even irregular,
nonicosahedral structures.

\subsection{Icosahedral Structures}

There are two different classes of icosahedral clusters: the ones with
a small particle in the center of the cage and those built around a
large particle. We will call these structures S-ICO and L-ICO
respectively. On fig. 2A,B,C,D the S-ICO peaks are marked with an
oval, and L-ICO peaks are signed with a rectangle.  First we note that
for the three lowest $\alpha$'s S-ICO's are present at low energy for
all concentrations, while for $\alpha$=1.6 they can be found only for
$\eta>6$. This observation can be summarized by saying that whenever
the energy gap is present, all states below the gap are S-ICO's
(although the presence of a gap for $\alpha$=1.4 is not so obvious at
small concentration).  L-ICO's are present only for low $\eta$'s,
intermixed with non-icosahedral states: their energy grows rapidly
with $\eta$, so that they can never be found for $\eta>5$ in the energy
range we considered: the presence and the importance of this kind of
structures is therefore very marginal. Figure~\ref{icol} summarize this
aspect: the lowest L-ICO configurations are plotted as function of the
concentration.\\
  It is also interesting to note the number of different
S-ICO configurations has a maximum for intermediate concentrations,
and it decreases simmetrically when $\eta$ gets larger or smaller. The
number of different icosahedral configurations built around a small
particle can be predicted. We can state the problem in the following
way : in how many ways can the 13 vertices of an icosahedron be
decorated with $\eta$ S-atoms and $13-\eta$ L-atoms ? if no vertices are
equivalent the answer is trivial: the number of different decorations
will be given by the binomial coefficient. But this is not the case of
the regular icosahedral cage: in this case the problem is to calculate
the number of different isomers (or different polytypes in
mathematical terminology). As an example consider the case of
$\eta$=1. In this case we have two different indipendent arrangements:
as first choice we can put the S particle in the center of the cage
and the L particles on vertices of the icosahedron. The second
possibility is to put an L particle in the center: for the S particle
there are therefore 12 topologically equivalent vertices on the
external cage: the resulting isomer is therefore 12 times
degenerated. The process of deriving all the isomers and the relative
degenerations for all $\eta$'s is not a simple task. The number of
different isomers has been recently calculated by Theo and coworkers
\cite{pic} with an application of the Polya's enumeration theorem. In
table 1 we report, as a function of $\eta$, the number of isomers and
the number of different S-ICO's found for each value of $\alpha$. For
the lowest two values of $\alpha$ the number of S-ICO's coincides with
the expected number of isomers: peaks at different energies below the
gap are due to different arrangements of L and S-atoms on an
icosahedral structure. For $\alpha$=1.4 the ratio between found S-ICO's
and expected isomers drops for values of $\eta$ smaller than 7, as
shown in fig.~\ref{figx}. An even more dramatic transition is shown for
$\alpha$=1.6, where the ratio drops abruptly from one to zero when
$\eta$ decreases from 9 to 6, showing the impossibility to build
icosahedra for such radii-ratio when the s-particles are minority.\\
Interesting observations can arise from the fact that for high values
of concentration almost all the isomers are present. We checked to
find correlations between the relative positions of L-atoms
(considered as impurites in a cluster formed by S-atoms) and the
energy or the distortion of the cluster. Our results do not allow a
coherent interpretation of all data, but we were able to detect some
regularities \cite{coz95}. Let us first focus on $\eta>9$, where we find
icosahedral
isomers for all the values of $\alpha$. For all these structures a
general rule is valid: the energy of the isomers is strictly
correlated with the disposition of the impurities on the cluster. As a
matter of fact in the lowest minima we find structures where the
L impurities tend to occupy contiguous positions on the
external shell, and the energy grows when the atoms are placed in non
adjacent positions. A very clear example is given in the case of two
large atoms ($\eta$=11 and 3 different isomers): the lowest
configuration is the configuration with impurities placed as first
neighbours, the second with L-atoms as second neighbours and the
highest one with L-atoms as third neighbours. This situation is the
same for all the values of $\alpha$. Decreasing $\eta$ this rule is
violated: for $\eta$=9 it remains valid for $\alpha$=1.33 and
$\alpha$=1.25 while for $\eta$=8 only the lowest value of $\alpha$ is
consistent with this picture. Lowering $\eta$ again no regularities are
found.  Along this line one can expect that the structures of the
icosahedra have a high degree of distorsion if the L-atoms are close,
while the distorsion is small if they are placed in a simmetric
way. All the observations of distorsion (based on the measure of the
third order invariant) confirm this idea so we can say that the most
distorded clusters correspond to the lowest minima. We have to stress
that this picture is valid only as long as the L-atoms can be
considered as a perturbation on a icosahedron of S-atoms, i.e. for
large $\eta$'s. A similar rule might be valid also for the case in
which S-atoms perturb an icosahedron of L-atoms, but since L-ICO's are
very expensive in energy we do not have samples to verify this idea.
With regard to the role of $\alpha$, it's easy to guess that the
distorsion grows with $\alpha$: our results confirm this conjecture for
all the values of concentrations.

\subsection{The case $\alpha=1.6$ and $\eta=1$}
\label{al}
As we anticipated the case $\alpha$=1.6 and $\eta$=1 is singular: it
is therefore interesting to discuss this case in detail, with the help
of 3-D images. The lowest energy is much higher than that for
$\eta>1$, and the frequency plot shows two twin peaks. However these
two peaks are degenerate, in the sense that each is formed by two
different structures very close in energy. The four configurations
occurr with the same frequency.  Figures~\ref{figxx}A and
{}~\ref{figxx}B present the two structures forming lowest energy peak:
the clusters have the same basic structure: a cage formed by ten large
atoms sorrounds a small atom at the center, while the other two large
atoms are on a second shell. The two different states are due to a
different accomodation of these external atoms.  Figures ~\ref{figxx}C
and ~\ref{figxx}D show the two clusters belonging to the second peak.
It's worth noting that the structures are completely different in that
case: the configuration at higher energy is an icosahedron with a
large particle in the center while the lower one is formed by the
small atom in the center surrounded by only 8 atoms: four atoms are on
the second shell of the cluster.  We observe none of the four
configuration is an S-ICO: the reason of its singularity lies in part
on this simple observation. In all cases for $\alpha \leq$ 1.4 the
lowest energy structures are S-ICO's. The same is true for
$\alpha$=1.6 and $\eta\geq 6$. For $\alpha$=1.6 and $1\leq \eta \leq5$
it is possible to construct structures with a droplet of two or three
S-atoms at the center (like in fig.~\ref{fig5} and ~\ref{fig6}),
allowing all the L-atoms to be embedded in a first shell around the
droplets. However when the size ratio is so large and only one S-atom
is available, putting the small atom at the center forces two large
particles on a second shell: these particles have a small coordination
and therefore the energy of the cluster is high.

\section{ Discussion}
\subsection{Interpretation of the Results}
In a monoatomic system, the best arrangement of 13 atoms is the
icosahedron. Our work is aimed to understand if and how this
arrangement changes when a 13 atom cluster is composed of two
different kinds of atoms. It is useful to remind that an optimal
icosahedral configuration is achieved when the radius of the external
particles is 1.06 times larger than the radius of the central
particle: in an icosahedral cluster made of equal spheres, the spheres
on the shell are therefore a bit loose.  Let us first focus on the lower
$\eta$ values: the $\eta$ small atoms can be seen as a perturbations
of a cluster made of L particles. Such perturbation can have opposite
effects, depending on whether an S-atom occupies the center of the
cluster or not.  We consider first the case $\eta$=1 in which the
icosahedron has an L particle at the center, while on the vertices
there are 11 L-atoms and 1 S-atom. In the monoatomic cluster the
particles on the shell are a bit loose: by shrinking the size of an
external particle frustration and energy are increased. When $\eta$
grows, so will do frustration and energy. These considerations can
explain what we observe: L-ICO clusters are found only for very small
$\eta$'s, and their energy grows rapidly with $\eta$. The role of the
$\alpha$ parameter is also easy to understand: the larger the
difference between the two component in the cluster, the faster the
growth in energy.  Let us now consider the case $\eta$=1 with the small
atom as the central particle of the cluster: in such configuration the
external particles are more squeezed than in the monatomic case. A
value of $\alpha$ larger than 1 but smaller than 1.06 helps removing
the frustration typical of monoatomic systems. For larger values of
$\alpha$ frustration is induced by the fact that the small particle at
the center is unable to ``fill'' the space in the cage created by the
external particles. Our results show that for a value of $\alpha$ as
large as 1.25 the energy decreases, which means that the overall
frustration is diminuished, while larger values of $\alpha$ induce a
growth in energy. Increasing $\eta$, the new S-atoms must substitute
some L-atoms on the external shell, which by shrinking decreases the
frustration of the central atom: energy therefore drops. S-ICO's are
therefore favoured in most cases, with a notable exception occurring
when $\alpha$ is too large and $\eta$ is small. In such case, less
than 12 particles are needed to surround the central atom: the
remaining L-atoms end up in a second shell, where their coordination
is low and therefore the energy of the cluster grows. This case was
discussed in detail in section \ref{al}.  The transition
from icosahedral structures to non icosahedral ones is clearly shown
from the shrinking energy gap between the S-ICO's and other non
icosahedral configuration with impurities at the center of the
cage. The non icosahedral minima are favoured as the gap in energy
becomes smaller and are dominant starting from $\alpha$=1.33.\\
  Let us
now consider the high $\eta$ values. Similar but opposite
considerations hold. At $\eta$=12 only one L-atom is present: if it
was at the center, frustration of the S-atoms on the shell would
increase. In fact, we never find L-ICO's in this range. The other case
(the S-atom as central particle) is the dominant one. The presence of
L-atoms on the shell increases the density and moves the cluster
toward an energetically more convenient configuration. By deacresing
$\eta$, this process continues: S-ICO's are stable in the whole
high-$\eta$ region. \\
 At the light of these considerations it is
interesting to reconsider figures 3 and \ref{gap}. As far as the three
lowest values of $\alpha$ are concerned, figure 3 shows how the effect
of impurities has a strong dependence on the choice of $\alpha$ in
the case of small $\eta$'s (i.e. a few impurities of type S), while
when the impurities are of type L (large $\eta$'s) the behavior does
not strongly depend on $\alpha$. The curve for $\alpha$=1.6 instead
shows two different minima, indicating that at small $\eta$'s
icosahedra are again the dominant structure, while at large $\eta$'s
different kinds of structures play the dominant role.  Figure \ref{gap}
shows how the insertion of small particles changes the monoatomic
case: the first substitution of an S-atom on the cluster lowers the
gap between icosahedral and non-icosahedral clusters. This gap tends
then to grow toward the monoatomic value when more S-atoms are
inserted. The parameter $\alpha$ changes the speed in reaching the limit of the
monoatomic case. In the case $\alpha$=1.6 this behaviour is found only
at high $\eta$'s because the absence of icosahedra elsewhere.

\subsection{Comparison with Extended Systems}
We focus now our attention on a set of papers treating the analysis of
icosahedral order in extended binary systems. Our aim is to see what
degree of agreement exists between the behaviour we described and the
presence or absence of icosahedral order detected in the extended
system discussed in these works. In table II we report a certain
number of papers dealing with the problem of icosahedral order. From
the data of the simulations of each of these works we have extrapolated
our two parameters $\eta$ and $\alpha$. It is clear that the right
comparison with the parameter $\alpha$ is possible only if the
interactive potential is a LJ and if the mixed interaction is treated
as in our work. Anyway for sake of completness we reported in the
table also studies performed on different systems: in this case the
definition of the $\alpha$ has to be done carefully.\\
  Let us first
discuss the system which are strictly comparable with our results:
LJ systems. We first concentrate on the work of J\'onsson and Andersen
\cite{jon88} and of Shumway {\it et al. }\cite{jon95}, in which both groups
reported a considerable amount of icosahedral order in their samples.
The parameters of two simulations are equal and correspond to
$\alpha$=1.25 and $\eta$=9.6. Thank to the courtesy of Hannes
J\'onsson that provide us some configurations we have analyzed
carefully some samples from both these works in order to see in more
detail what kind of icosahedra are present: we detected in all the
samples only S-ICO types of icosahedra. The characteristics of the
icosahedra (structures and concentrations) are in perfect agreement
with the results discussed previously \cite{coz94}.  Our results are
also in excellent agreement with the work by Dasgupta {\it et al.
}\cite{das91}. In their paper they reported absence of long range
icosahedral correlation, which at the light of our study is no
surprise, since at the values of concentration and radii chosen by
them ($\alpha$=1.6, $\eta$=6.5) there can hardly be an icosahedron!
Their result should therefore not be taken as a general indication of
the absence of icosahedral order in LJ systems.\\
Finally the
comparison between our results and the ones by Ernst {\it et al.
}\cite{gre} gives some interesting notes. Let us discuss what kind of
icosahedra should be expected to find in an extended system like the
one simulated by the authors at the light of our results. In the
region defined by their parameters ($\alpha$=1.25, $\eta$=2.6) we
found icosahedral cluster with both S-ICO and L-ICO clusters, with a
strong predominace of the former, due to energetic reasons. When we
cool down such extended system we would not expect local minima of
type L-ICO (due to their high cost in energy). We would instead expect
to find a certain number of S-ICO confgurations, since these are
energetically favoured: however this number could not be very high,
because the small concentration of S-atoms which are needed as seeds
of such a kind of icosahedra. This picture fits very well with the
results of the work performed by Ernst {\it et al. }. They in fact say
{\it ..the data for $Q_{l}$ suggest icosahedral simmetry ...  and the
average in $Q_{l}$ is less for a pure icosahedron}. A similar result
is also obtained by J\'onnson \cite{jon94} on an equivalent system
(same values of $\alpha$ and $\eta$), in which again only relatively
few icosahedral shells are found, and all of them have S-atoms at the
center.\\ Let us now discuss some works with different potential. One
of the most interesting paper we found is the one of Qi and Wang
\cite{qii} in which the authors simulated an $Mg_3Ca_7$ system finding
a strong evidence of icosahedral order. they reported a complete
cluster analysis identified by the mean of CNA. Having checked that
their definition of the polyhedra is equivalent to our Voronoi
construction we then can compare our polyhedra with theirs. The main
problem in this case is to define the parameter $\alpha$, since the
authors did not report the core radius of the pair potential used. We
estimate $\alpha$ to be equal to 1.33 where Ca is the L-atom and Mg
the small one (this value of the ratio is the same suggested by Nelson
\cite{nel89} for the same system modelled by hard sphere. It is also
very close to the ratio of the atomic radii of the two elements.).
Also in this case let us imagine what kind of structures would be
expected to see in such system based on our findings: with $\alpha$=1.33
and $\eta$=4 we are in zone with dominant S-ICO clusters but with a
few other minima of some importance in the gap. So what we expect to
see is some icosahedral structures (all with small atoms at the
centre) and some other significant minima. This picture agrees with
the results by Qi and Wang partially: the most frequented clusters are
coordinate 12, and the structures are icosahedral and (0,2,8,2,0)
polyhedra. A rough extimation of the frequency of these polyhedra is
qualitatively similar with our frequency of visits. A great majority
of the icosahedral cluster has a small atom (Mg) at the center: the
surprise however is the existence of some L-ICO structures. The
relative absence of low coordination polyhedra is probably due to the
difference between an extended system and an isolated cluster: however
all the low coordination polyhedra they detected are present in our
configurations.\\ Finally we compare our result with two works
performed on hard sphere systems. In this case the natural choiche of
the parameter $\alpha$ is given by the ratio of the radii of the two
different spheres. Our results fit well with the result of Clarke and
J\'onnson \cite{cla93b} in which they found icosahedral order in a
binary system of hard sphere with parameter $\alpha$=1.25 and
$\eta$=9.6. In the work of Clark and Wiley \cite{cla88} in which they
explore several concentrations and $\alpha$ values the results are more
difficult to compare directly because the authors do not present
extended results for all the points explored and they limit themselves
to general considerations concluding that icosahedral order is
neglible in all their systems. However extrapolating the data from the
figures presented we can detect a qualitative agreement also in this
case.\\ As a final remark we can say that our results show a
substantial agreement with all the results presented in literature and
can therefore be used as a good indication for predicting the
existence of icosahedra (and maybe of icosahedral order) for given
size-ratios and concentrations in binary systems.
\section{Conclusion}
We studied 13-atom clusters of binary mixtures at four radii-ratios
$\alpha$ and for all possible concentration $\eta$ of small
particles. We found that icosahedral structures are important for
small $\alpha$ and become less relevant when $\alpha$ grows: for large
values of $\alpha$ icosahedral structures play a role only when $\eta$
is large. A general interpretation of this behaviour is given in in
term of impurities on the clusters. Finally our results can be
successfully compared with the claims of presence or absence of
icosahedra in quenched extended binary systems, therefore allowing a
better uderstanding of the contrasting results which appear in
literature.

\section {Acknowledgments}
One of us (SC) spent part of the time during this work at the
Department of Chemistry of the University of Washington. We are
grateful to Hannes J\'onsson for hospitality and for having provided the
samples to analyze and for the many useful discussions. This work has
been supported in part by the italian CNR Progetto Finalizzato Sistemi
Informatici e Calcolo Parallelo (U.O.Reatto).

%
%
\begin{figure}
\caption{The set of the energy spectra for all the points in our space
 parameter. Figures 1A, 1B, 1C 1D represent respectivetely the spectra
 for $\alpha$=1.25,1.33, 1.4, and 1.6. Within the same $\alpha$ all the
 12 different concentrations are plotted starting with $\eta$=1 (left)
 and ending with $\eta$=12 (right). The energy scale (the same for all
 the plots) is the absolute scale of energy and every straight line
 represent a different minimum. All the minima reported here are
 detected at least twice in our simulations.  }
\label{fig1}
\end{figure}

\begin{figure}
\caption{The histograms of the percentage of visits to the minima. As
in figure 1 each of four different figures is for the four different
values of $\alpha$ and within the same $\alpha$ all the 12 different
values of concentrations are plotted. For a better representation
every different value of concentrations is shifted up by a number
proportional to $\eta$: the lowest curve is therefore for the lowest
concentration ($\eta$=1) while the upper curve is for $\eta$=12. Every
peak of the curve is generally formed by more than a single minimum
because the relative large bin used in the histograms. All the peaks
signed by an oval are due to icosahedral structures with a small atom
at the center of the cage and the contribution to those peaks comes
completely from icosahedral minima. On the curves we also indicated
the positions of the icosahedral minima (squares) with a large atom at
the center. In those cases, conversely, the contribution given by the
icosahedral structures is not dominant at all.}
\label{fig2}
\end{figure}

\begin{figure}
\caption{ The lowest minimum in the spectra plotted as a function of
the concentration for the four values of $\alpha$. The energy is in
LJ units. The three lower values of $\alpha$ has a similar parabolic
shape with the same minumum at $\eta$=7 and all the minima have an
icosahedral cage. While the 1.6 case has a double well-shape and the
icosahedral structure are limited only for $\eta>7$.  }
\label{fig3}
\end{figure}

\begin{figure}
\caption{ The energy gap between the highest minimum in energy of the
 S-ICO's structures and the
first non icosahedral cluster as a function of the
concentrations. The four lines indicate the four different values of
$\alpha$. The 1.6 curve is not over all the range because there are
icosahedral structure for $\eta < 6$. Energy is given in LJ units.}
\label{gap}
\end{figure}

\begin{figure}
\caption{ In this set of four figures we plot the percentage of the
clusters with a Voronoi polyhedra with 12 faces (indicates with 12) as
a function of concentration and the percentage of the remaining
clusters with less than 12 faces (indicates with others). As usual A B
C D refer to $\alpha$=1.25, 1.33, 1.4 and 1.6. For the greater three
value of $\alpha$ we plot also the major component of the other curve:
for $\alpha$=1.33 and 1.4 the percentage of clusters with the 10-faces
Voronoi cage, while for $\alpha$=1.6 the dominant component is given by
polyhedra with 9 faces.}
\label{fig4}
\end{figure}

\begin{figure}
\caption{A typical cluster with a Voronoi cage of 10 faces for $\eta$=3
and $\alpha$=1.4. The cage in this case has 2 quadrangolar faces and 8
pentagonal faces. The two external particles are indicated in light
gray. The central particle here is a small one.
Note that the 3 small atoms are all first
neighbours.}
\label{fig5}
\end{figure}

\begin{figure}
\caption{The absolute minimum energy cluster for the case $\alpha$=1.6
and $\eta$=2. The two small particles are inside a cage formed by the
large particles. The peculariaty of this structure is that each of the
two small particles is at the center of a Voronoi cage of 9
faces. The particles belonging to a such a structure are in the figure
in light gray).In dark gray are represented the three particles on the
second shell of such a cage.  }
\label{fig6}
\end{figure}

\begin{figure}
\caption{ The ratio between the number of different icosahedral minima
detected for $\alpha$=1.4 and 1.6 and the number of isomers as a
function of concentrations. All the icosahedral minima are of
S-ICO-type.}
\label{figx}
\end{figure}

\begin{figure}
\caption{The energy of the lowest L-ICO structures as a function of
the concentration $\eta$ . The four symbols are for the different
values of $\alpha$. The rapid increasing in energy as $\eta$ grows
limits the range of presence of such structures at very low
concentrations.}
\label{icol}
\end{figure}

\begin{figure}
\caption{ The four different structure of the lowest minima for the
case $\alpha$=1.6 and $\eta$=1. Fig A and B are the lowest two, with the
same Voronoi cage (0,2,8,0,0): in this case the small differences in
energy is given by the different positions on the second shell of the
two remaining particles. Figure C is the perfect icosahedron with the
small particle on the external shell and with a large particle at the
center while figure D shows the last minimum in which the small
particle is sourrounded by only few atoms L, and the remaining atoms
take place on the second shell.  }
\label{figxx}
\end{figure}

%
%
\begin{table}
\caption{ The table presents the number of different isomers for an
icosahedron having a particle S at the center, and $13 - \eta$ atoms of the
second kind (L) distributed on external vertices. The number of
isomers is then compared with the number of different icosahedral
minima (with S particle at the center) detected for all the values of
$\alpha$.}
\label{tab1}
\begin{tabular}{cccccccr}
$\eta$ & isomers & $\alpha$=1.6 & $\alpha$=1.4 & $\alpha$=1.33 &
$\alpha$=1.25 \\ \hline
1&1&0&1&1&1\\
2&1&0&1&1&1\\
3&3&0&1&3&3\\
4&5&0&3&5&5&\\
5&10&0&7&10&10\\
6&12&0&9&12&12&\\
7&18&2&17&18&18\\
8&12&5&12&12&12&\\
9&10&9&10&10&10\\
10&5&5&5&5&5&\\
11&3&3&3&3&3\\
12&1&1&1&1&1\\
\end{tabular}
\end{table}

\begin{table}
\caption{A collection of papers dealing with the problem of
icosahedral order in binary systems. Column one gives the authors,
column two specifies the interatomic potential used in
simulations. The percentage of S particles used by the authors is
given in column 3, and in column 4 as $\eta$, to
facilitate the comparison with the present work. Column 5 indicates
the value of the size ratio ($\alpha$), and column 6 gives an
indication about icosahedra detected in the samples. }
\label{tab2}
\begin{tabular}{llllllcr}\\
Authors & potential& \% of S-atoms & $\eta$ & $\alpha$ &
icosahedra? \\ \hline
J\'onsson and Andersen (ref. \cite{jon88}) & LJ &
80\% & 9.6 & 1.25 & YES \\
Shumway {\it et al. }(ref \cite{jon95})& LJ & 80\% & 9.6 & 1.25 &YES \\
Dasgupta {\it et al. }(ref \cite{das91}) & LJ & 50\% & 6.5 & 1.6 & NO \\
Ernst {\it et al.} (ref. \cite{gre}) & LJ& 20\% & 2.6 &
1.25 & Suggested \\
Qi and Wang (ref \cite{qii})& Mg3Ca7 & 70\% & 3.9 & 1.33 &YES \\
Clarke and Wiley (ref \cite{cla88})& hard-spheres & various & various & 1.1 -
1.5 & Negligible \\
Clarke and
J\'onsson (ref \cite{cla93b})& hard-spheres & 80\%& 9.6 & 1.25 & YES\\
\end{tabular}
\end{table}
\end{document}